# Approximating the Region of Multi-Task Coordination via the Optimal Lyapunov-Like Barrier Function

Dongkun Han, Lixing Huang, and Dimitra Panagou

*Abstract*—We consider the multi-task coordination problem for multi-agent systems under the following objectives: 1. collision avoidance; 2. connectivity maintenance; 3. convergence to desired destinations. The paper focuses on the safety guaranteed region of multi-task coordination (SG-RMTC), i.e., the set of initial states from which all trajectories converge to the desired configuration, while at the same time achieve the multi-task coordination and avoid unsafe sets. In contrast to estimating the domain of attraction via Lyapunov functions, the main underlying idea is to employ the sublevel sets of Lyapunov-like barrier functions to approximate the SG-RMTC. Rather than using fixed Lyapunov-like barrier functions, a systematic way is proposed to search an optimal Lyapunov-like barrier function such that the under-estimate of SG-RMTC is maximized. Numerical examples illustrate the effectiveness of the proposed method.

## I. INTRODUCTION

Assessing the stability properties of an equilibrium point is of fundamental significance in control and dynamical systems theory. For asymptotically stable equilibrium points, one long-standing and in practice exceedingly difficulty problem is the estimation of the region of attraction, i.e., of the set of initial states from which all trajectories converge to the equilibrium point.

In addition, with the rapid recent developments in communication and sensing technologies, ubiquity of multi-agent systems has spurred great research interest in areas such as multi-robot path planning, surveillance (for more applications, refer to surveys [1], [2] and books [3], [4]). Apart from stability of the concerned equilibrium points, efficient coordination of multi-agent systems typically requires connectivity maintenance and collision avoidance amongst agents. Thus, the following questions arise naturally: Is it possible to compute the region of coordination for multi-agent systems while guaranteeing convergence, collision avoidance and connectivity maintenance? How can we estimate the region of multi-task coordination? To the best of our knowledge, these issues have not been addressed yet and still remain challenging.

Dongkun Han is with the Department of Aerospace Engineering, the University of Michigan and the Department of Mechanical and Automation Engineering, the Chinese University of Hong Kong. E-mail: dongkunh@umich.edu. Lixing Huang and Dimitra Panagou are with the Department of Aerospace Engineering, the University of Michigan. E-mail: {lixhuang,dpanagou}@umich.edu.

This work was sponsored by the Automotive Research Center (ARC) in accordance with Cooperative Agreement W56HZV-14-2-0001 U.S. Army TARDEC in Warren, MI, USA, and an Early Career Faculty Grant from NASA's Space Technology Research Grants Program. Toyota Research Institute ("TRI") provided funds to assist the authors with their research but this article solely reflects the opinions and conclusions of its authors and not TRI or any other Toyota entity.

In order to answer these questions, let us first review the methods for estimating the region of attraction of isolated dynamical systems. The sublevel set of Lyapunov function is proven to be a useful way, in which different types of Lyapunov functions are employed; from the simplest form, i.e., quadratic Lyapunov functions, to more complicated forms, such as pointwise maximum Lyapunov functions or rational polynomial Lyapunov functions (see [5] and references therein). Nevertheless, the sublevel set of Lyapunov functions cannot in principle guarantee cooperative objectives such as collision avoidance and connectivity maintenance.

To achieve multi-task coordination, Lyapunov-like barrier functions are able to encode the constraints of each agent, and provide simple but effective, gradient-based control strategies. According to different objectives, various elegant Lyapunov-like scalar functions are proposed, including potential functions [6], navigation functions [7], harmonic functions [8], barrier functions [9], and avoidance functions [10]. However, the Lyapunov-like functions are usually selected with fixed forms, which result in conservative results when it comes to the estimation problem of SG-RMTC. In [11], a compositional barrier function is proposed by using logical operators, but the barrier functions are also fixed for the corresponding objectives. In [12], a barrier certificate is constructed using Sum-of-Squares decomposition. However, this method is merely used for safety verification, without guaranteeing the convergence of trajectories to desired equilibrium points, thus not applicable to multi-task coordination.

Motivated by aforementioned results, and based on our previous work [13], [14] that uses fixed Lyapunov-like barrier functions, this paper proposes a systematic way to generate a feasible Lyapunov-like barrier function, and gives a method to maximize the largest estimate of SG-RMTC via the optimal Lyapunov-like barrier function, which provides a larger stability margin compared to the fixed ones. The novelties of this paper lie in the following aspects:

- Based on the real Positivestellensatz, the estimation problem of SG-RMTC boils down to a Sum-of-Squares programming. By employing the Square Matrix Representation technique, a lower bound of the largest estimate of the SG-RMTC can be computed by solving a generalized eigenvalue problem.
- Different from other work that uses fixed Lyapunov-like barrier functions [15], [16], a systematic way is proposed for searching feasible polynomial Lyapunov-like barrier functions. In addition, a strategy is given for pursuing the optimal Lyapunov-like barrier function such that the estimate of SG-RMTC can be maximized.

## II. PRELIMINARIES

Notations: $\mathbb{N}, \mathbb{R}$: natural and real number sets; $\mathbb{R}_+$: positive real number set; $A^T$: transpose of $A$; $A > 0$ ($A \geq 0$): symmetric positive definite (semidefinite) matrix $A$; $A \otimes B$: Kronecker product of matrices $A$ and $B$; $\text{diag}(a)$: a square diagonal matrix with the elements of vector $a$ on the main diagonal; $\|a\|$: Euclidean norm or $l_2$ norm of vector $a$; $\deg(f)$: degree of polynomial function $f$; $(*)^T A B$ in a form of Square Matrix Representation: $B^T A B$. Let $\mathcal{P}$ be the set of polynomials and $\mathcal{P}^{n \times m}$ be the set of matrix polynomials with dimension $n \times m$. A polynomial $p(x) \in \mathcal{P}$ is nonnegative if $p(x) \geq 0$ for all $x \in \mathbb{R}^n$. A useful way of establishing $p(x) \geq 0$ consists of checking whether $p(x)$ can be described as a sum of squares of polynomials (SOS), i.e., $p(x) = \sum_{i=1}^{k} p_i(x)^2$ for some $p_1, \ldots, p_k \in \mathcal{P}$. The set of SOS polynomials is denoted by $\mathcal{P}^{\text{SOS}}$.

### A. Model Formulation

Each agent is modeled by the double-integrator model as follows:
$$\begin{array}{rcl} \dot{x}_i(t) & = & \rho_i(t) \\ \dot{\rho}_i(t) & = & u_i(t), \quad i \in \mathcal{N}, \end{array} \quad (1)$$

where $\mathcal{N} = \{1, \ldots, N\}$, $x_i(t) \in \mathbb{R}^n$ denotes the position state, $\rho_i(t) \in \mathbb{R}^n$ denotes the velocity state, and $u_i(t) \in \mathbb{R}^n$ denotes the control input on $i$-th agent. In the sequel, we will omit the arguments $t$ and $x$ of functions whenever possible for the brevity of notations.

A *weighted undirected dynamic graph* $\mathcal{G}(t) = (\mathcal{A}, \mathcal{E}(t), G)$ is used to describe a network of multi-agents, with the set of nodes $\mathcal{A} = \{A_1, \ldots, A_N\}$, the set of undirected edges $\mathcal{E}(t) = \{(A_i, A_j) | A_i, A_j \in \mathcal{A}\}$, and the *weighted adjacency matrix* $G = (G_{ij})_{N \times N}$. Fig. 1 shows the model of agents and the switching law of edges $\mathcal{E}(t)$.

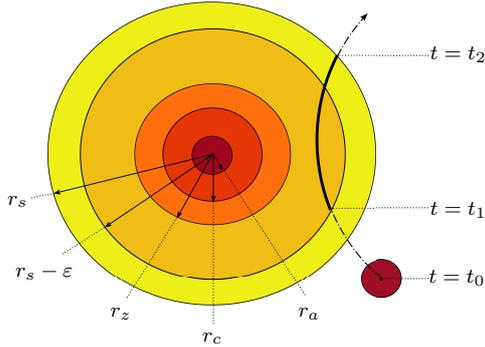

Fig. 1. The agent model and changing rules of edges: $r_a$ denotes the radius of each agent; $r_c$ is the radius of collision avoidance area; $r_z$ denotes the radius of area that the control with collision avoidance objective is active; $r_s$ denotes the radius of sensing area; constant $\epsilon \in [0, r_s - r_z]$ is a distance parameter for the hysteresis in adding new edges. The solid line for $t \in (t_1, t_2)$ shows the part of trajectory when there is an edge between these two agents.

A graph $\mathcal{G}(t)$ is *connected* at time $t$ if there is a path between any pair of distinct nodes $A_i$ and $A_j$ in $\mathcal{G}(t)$. The *Laplacian matrix* is given as $L(t) = \Delta(t) - G(t)$ where $\Delta(t) = \text{diag}(\sum_{j=1}^{N} G_{ij}(t))$. A relationship between $L(t)$ and the connectedness of $\mathcal{G}(t)$ is given in [4]:

*Lemma 1:* Let $\lambda_1(L(t)) \leq \lambda_2(L(t)) \leq \cdots \leq \lambda_N(L(t))$ be the ordered eigenvalues of $L(t)$. Then, $1_N$ is an eigenvector of $L(t)$ with the corresponding eigenvalue $\lambda_1(L(t)) = 0$. Moreover, $\lambda_2(L(t)) > 0$ if and only if $\mathcal{G}(t)$ is connected. □

### B. Problem Formulation

The distributed controller of agent $i$ depends on the local information of agent $i$, i.e., relative distances, relative velocities, and the coupling weights of communications. Specifically,
$$u_i = \sum_{j \in \mathcal{N}_i^s(t)} f\Big(x_i(t) - x_j(t), \ \rho_i(t) - \rho_j(t), \ G_{ij}(t)\Big), \quad (2)$$

where $\mathcal{N}_i^s(t) = \{j | (A_i, A_j) \in \mathcal{E}(t)\}$ is the neighborhood set of agent $i$ (in the sensing range of agent $i$). System (1) can be rewritten as:
$$\dot{q} = g(q), \quad (3)$$

by introducing $y_i = x_i - \tau_i$, $\varrho_i = \rho_i - \rho^*$, $q_i = (y_i, \varrho_i)^T$, $q = (q_1^T, q_2^T, \ldots, q_N^T)^T$, where $\tau_i$ and $\rho^*$ are the ideal displacement and the desired velocity of agent $i$ in the desired formation configuration, respectively.

Consider system (3), $\mathcal{U} \in \mathbb{R}^{2N}$ is an undesired set, and the origin $0_{2N}$ is an equilibrium point of the system. Let $V(q) : \mathbb{R}^{2N} \to \mathbb{R}$ be a continuously differentiable function on $q$ such that: 1) $V(0_{2N}) = 0$ and $V(q) > 0$ in $\mathbb{R}^{2N}/\{0_{2N}\}$; 2) $\dot{V}(q) < 0$ in $\mathbb{R}^{2N}/\{0_{2N}\}$; 3) $V(q) = \infty$, for all $q \in \mathcal{U}$. Then, $\bar{q} = 0_{2N}$ is asymptotically stable, and $V(q)$ is called a *Lyapunov barrier function*. In addition, if condition 3) is changed to the condition of Barbashin-Krasovskii-LaSalle invariance principle, i.e., only the trivial solution $\bar{q} = 0_{2N}$ can stay identically in $\{q \in \mathbb{R}^{2N} | \dot{V}(q) = 0\}$, then $\bar{q} = 0_{2N}$ is asymptotically stable, and $V(q)$ is called a *Lyapunov-like barrier function*.

*Definition 1:* The *region of multi-task coordination* (RMTC) is expressed as
$$\mathcal{R} = \Big\{ q(0) \in \mathbb{R}^{2N} : \lim_{t \to +\infty} \chi(t; q(0)) = 0_{2N},$$
$$\mathcal{G}(t) \text{ is connected}, \ \|x_i(t) - x_j(t)\| > d_s, \forall t \geq t_0 \Big\},$$

where $\chi$ is the solution of system (3), $d_s$ denotes a user-defined safety distance for collision avoidance. □

In many practical implementations, an unsafe set is usually given for the situations where the system is at a great risk. The unsafe set in this paper is defined by polynomials as:
$$\Omega(t) = \Big\{ q(t) \in \mathbb{R}^{2N} : \omega_i(q) > 0, \ i = 1, \ldots, h. \Big\}, \quad (4)$$

and the safe set $\Omega^c(t)$ is the complement set of $\Omega(t)$. Based on this, we propose the set of interest as follows:

*Definition 2:* The *safety guaranteed region of multi-task coordination* (SG-RMTC) is described as
$$\mathcal{R}^{\text{SG}} = \Big\{ q(0) \in \mathbb{R}^{2N} : q(0) \in \mathcal{R}, \ q(t) \in \Omega^c(t), \ \forall t \geq t_0 \Big\}. \quad (5)$$

The sublevel set of Lyapunov-like function is used to estimate the SG-RMTC. Specifically, let $W(q)$ be a Lyapunov-like function of system (3) for the origin, which satisfies

$$W(0_{2N}) = 0, \ \forall q \in \mathbb{R}_0^{2N}: \ W(q) > 0, \ \lim_{\|x\| \to \infty} W(q) = \infty, \tag{6}$$

the time derivative of $W(q)$ along the trajectories of (1) is locally non-positive, and $0_{2N}$ is the only solution which can stay identically in $\{q| \ \dot{W}(q) = 0\}$ [17]. To this end, we introduce the sublevel set of $W(q)$ as

$$\mathcal{W}(c) = \left\{ q \in \mathbb{R}^{2N}: W(q) \leq c \right\}, \tag{7}$$

where $c \in \mathbb{R}^+$. For system (3), $\mathcal{W}$ is an estimate of $\mathcal{R}$ if

$$\forall q \in \mathcal{W}(c): \ \dot{W}(q) \leq 0, \tag{8}$$

and the time derivative of $W(q)$ along the trajectories of (1) is locally non-positive, and $0_{2N}$ is the only solution which can stay identically in $\{q| \ \dot{W}(q) = 0\}$. Let us propose the main problem we are concerned with:

*Problem 1:* Find a *polynomial* Lyapunov-like barrier function $W(q)$ and a positive scalar $c$ such that the estimate of the SG-RMTC is maximized under certain selected criteria, i.e., solving

$$\begin{aligned} \mu = \sup_{c, \ W} \ \zeta(\mathcal{W}(c)) \\ \text{s.t. } (6) - (8) \text{ hold,} \end{aligned} \tag{9}$$

where $\zeta$ is a measure of $\mathcal{W}(c)$ as a user-defined criteria, e.g., the volume of $\mathcal{W}(c)$. In addition, a gradient-based controller $u_i$ can be obtained in the form of (2) such that

1) $\lim_{t \to \infty} \|(x_i(t) - \tau_i) - (x_j(t) - \tau_j)\| = 0$, and $\lim_{t \to \infty} \|\rho_i(t) - \rho_j(t)\| = 0$, for all $i \in \mathcal{N}$ and $j \in \mathcal{N}_i^{\text{f}}$.
2) $\mathcal{G}(t)$ is connected, for all $t > t_0$, where $t_0$ is the initial time.
3) $\|x_i(t) - x_j(t)\| > d_s$, for all $t > t_0$, where $d_s$ denotes a user-selected safe distance for collision avoidance.
4) $q(t) \in \Omega^c$, for all $t \geq t_0$. $\square$

Some useful sets are introduced here: $\mathcal{N}_i^{\text{f}}$ is the neighborhood set to agent $i$ in the desired configuration, i.e., $\mathcal{N}_i^{\text{f}} = \{j| \ (A_i, A_j) \in \mathcal{E}^{\text{f}}, \ \|x_i - \tau_i - (x_j - \tau_j)\| = 0\}$, where $\tau_i$ is the ideal displacement of agent $i$ in the desired configuration, whose edge set is $\mathcal{E}^{\text{f}}$; We also define sets $\mathcal{N}_i^{\text{sf}}(t) = \{j| \ j \in \mathcal{N}_i^{\text{s}}(t), \ j \in \mathcal{N}_i^{\text{f}}\}$ and $\mathcal{N}_i^{\text{sz}}(t) = \{j| \ j \in \mathcal{N}_i^{\text{s}}(t), \ \|x_i - x_j\| < r_z\}$, which will be used in Section III.

For this problem, we assume that:

- Assumption 1: The desired configuration given by $\tau_i$ is achievable, i.e., $r_z \leq \|\tau_i - \tau_j\| \leq r_s - \varepsilon$, for all $i \in \mathcal{N}$, $j \in \mathcal{N}_i^{\text{f}}$. In other words, the desired distance between agent $i$ and agent $j \in \mathcal{N}_i^{\text{f}}$ is always between $r_s - \varepsilon$ and $r_z$.
- Assumption 2: The neighbor set of agent $i$ at time $t_0$ satisfies $\mathcal{N}_i^{\text{f}} \subseteq \mathcal{N}_i^{\text{s}}(t_0)$, which means that the desired topology is contained in the initial graph.
- Assumption 3: To achieve both objectives of collision avoidance and connectedness maintenance, we require $r_s - \|\tau_{ij}\| > d_s + \|\tau_{ij}\|$, for all $i, j \in \mathcal{N}$.

## III. MAIN RESULTS

### A. Controller Design with Local Connectivity Maintenance

In this paper, we use Lyapunov-like barrier functions to encode collision avoidance and connectedness maintenance. Other than using fixed Lyapunov-like barrier functions, this paper provides a systematic way to generate a feasible Lyapunov-like barrier function, from which a gradient-based controller can be obtained. For the brevity of expressions, let $\tau_{ij} = \tau_i - \tau_j$, $y_{ij} = y_i - y_j$, and $x_{ij} = x_i - x_j$.

For connectedness maintenance, from Assumption 2, the desired topology is contained in the initial graph. The main idea is to preserve the desired topology $\mathcal{E}^{\text{f}} \subseteq \mathcal{E}(t)$ such that the network is always connected for $t \geq t_0$. To do this, we would like to make the following condition satisfied: $\|x_{ij}\| < r_s$, for all $i \in \mathcal{N}$ and $j \in \mathcal{N}_i^{\text{sf}}(t)$ which holds if $r_s - \|\tau_{ij}\| - \|y_{ij}\| > 0$. Thus, the following barrier function $\Upsilon_{ij}^{\text{e}}(\|y_{ij}\|)$ is used with the constraints:

$$\begin{aligned} &\Upsilon_{ij}^{\text{e}}(\|y_{ij}\|) \geq 0, \ \Upsilon_{ij}^{\text{e}}(0) = 0, \ \Upsilon_{ij}^{\text{e}}(\hat{r}_s) = \mu_1, \\ &\frac{\partial \Upsilon_{ij}^{\text{e}}(\|y_{ij}\|)}{\partial(\|y_{ij}\|)} > 0, \ \forall 0 \leq \|y_{ij}\| \leq \hat{r}_s, \\ &\frac{\partial \Upsilon_{ij}^{\text{e}}(\|y_{ij}\|)}{\partial(\|y_{ij}\|)} \cdot \frac{1}{\|y_{ij}\|} > 0, \ \forall j \in \mathcal{N}_i^{\text{sf}}(t), \end{aligned} \tag{10}$$

where $\hat{r}_s = r_s - \|\tau_{ij}\|$, $\mathcal{N}_i^{\text{sf}}(t) = \{j| \ j \in \mathcal{N}_i^{\text{s}}(t), \ j \in \mathcal{N}_i^{\text{f}}\}$ defined in Sectioin II, $\mu_1$ is a positive scalar such that $\Upsilon_i^{\text{e}}$ is bounded when $\|y_{ij}\|$ tends to $\hat{r}_s$.

For collision avoidance, the basic idea is to keep the distance between any two agents $i$ and $j$ greater than a minimum user-defined safety distance $d_s > 2r_c$, where $r_c$ is given in Fig. 1. In other words, the condition is required that $\|x_{ij}\| > d_s$, which holds if $\|y_{ij}\| - d_s - \|\tau_{ij}\| > 0$. Thus, the following barrier function $\Upsilon_{ij}^{\text{c}}$ is introduced:

$$\begin{aligned} &\Upsilon_{ij}^{\text{c}}(\|y_{ij}\|) \geq 0, \ \Upsilon_{ij}^{\text{c}}(\hat{d}_s) = \mu_2, \\ &\frac{\partial \Upsilon_{ij}^{\text{c}}(\|y_{ij}\|)}{\partial(\|y_{ij}\|)} < 0, \ \forall \|y_{ij}\| \geq \hat{d}_s, \ \forall j \in \mathcal{N}_i^{\text{sz}}(t), \end{aligned} \tag{11}$$

where $\hat{d}_s = d_s + \|\tau_{ij}\|$, and $\mathcal{N}_i^{\text{sz}}(t) = \{j| \ j \in \mathcal{N}_i^{\text{s}}(t), \ \|x_{ij}\| < r_z\}$ introduced in Section II. $\mu_2$ is a positive scalar such that $\Upsilon_i^{\text{c}}$ is bounded when $\|y_{ij}\|$ tends to $\hat{d}_s$.

*Remark 1:* We assume $\mu_1$ and $\mu_2$ satisfying $\mu_1 > \mu_{\max}$ and $\mu_2 > \mu_{\max}$ with $\mu_{\max} := \frac{1}{2} \sum_{i=1}^{N} (\sum_{j \in \mathcal{N}_i^{\text{f}}} \Upsilon_{ij}^{\text{e}}(\|\hat{r}_s - \hat{\varepsilon}\|) + y_i(t_0)^T \sum_{j=1}^{N} G_{ij}(t_0) y_{ij}(t_0) + \rho_i(t_0)^T \rho_i(t_0)) + (N-1)N\Upsilon_{ij}^{\text{c}}(\|\hat{d}_s - \hat{\varepsilon}\|)$, where $0 < \hat{\varepsilon} < \min\{\frac{1}{2} d_s - r_c, \varepsilon\}$. The barrier function proposed in this paper is different than what is proposed in the existing relevant work [6], [9], [15], [16], [18]. In addition, collision avoidance [6], [9], [12], [15], [18], bounded control input [9], [12], [15], [16], and safety guaranteeing [6], [9], [15], [16], [18] are not considered. $\square$

For the brevity of notations, let us introduce $\Upsilon_i^{\text{e}} = \sum_{j \in \mathcal{N}_i^{\text{sf}}} \Upsilon_{ij}^{\text{e}}$, $\Upsilon_i^{\text{c}} = \sum_{j \in \mathcal{N}_i^{\text{sz}}} \Upsilon_{ij}^{\text{c}}$, $x = (x_1^T, x_2^T, \ldots, x_N^T)^T$, $\rho = (\rho_1^T, \rho_2^T, \ldots, \rho_N^T)^T$. A distributed controller is provided as follows:

$$u_i = -\alpha^{\text{e}} - \alpha^{\text{c}} - \beta^y - \beta^\rho, \tag{12}$$

where $\alpha^{\mathrm{e}} = \sum_{j \in \mathcal{N}_i^{\mathrm{sf}}(t)} \nabla_{y_i} \Upsilon_{ij}^{\mathrm{e}}(\|y_{ij}\|)$, $\alpha^{\mathrm{c}} = \sum_{j \in \mathcal{N}_i^{\mathrm{sz}}(t)} \nabla_{y_i} \Upsilon_{ij}^{\mathrm{c}}(\|y_{ij}\|)$, $\beta^y = \sum_{j \in \mathcal{N}_i^{\mathrm{s}}(t)} G_{ij}(t) y_{ij}$, $\beta^\rho = \sum_{j \in \mathcal{N}_i^{\mathrm{s}}(t)} G_{ij}(t) \rho_{ij}$, $G_{ij}$ is the $ij$-th entry of weighted adjacency matrix. The following result shows that under conditions (10) and (11), the multi-task coordination is guaranteed by the feasible gradient-based controller (12).

*Theorem 1:* If Assumption 1-3 holds, and $\mathcal{G}(t_0)$ is connected, then, under the controller (12), the following conditions hold for all $i \in \mathcal{N}$:

1) $\mathcal{G}(t)$ is connected for all $t \geq t_0$;
2) Collision avoidance is ensured for all $t \geq t_0$.
3) $\lim_{t \to \infty} \|\rho_i - \rho_j\| = 0$, for $j \in \mathcal{N}$;
4) $\lim_{t \to \infty} \|x_i(t) - \tau_i - (x_j(t) - \tau_j)\| = 0$, for $j \in \mathcal{N}_i^{\mathrm{f}}$.

*Proof*: For statement 1) and statement 2), we aim to show the concerned set is a forward invariant set, which implies the connectedness and collision avoidance. Specifically, we assume that the edge set $\mathcal{E}(t)$ changes at $t_l$, $l = 0, 1, 2, \ldots$. For each $[t_l, t_{l+1})$, $\mathcal{G}$ is fixed. Based on (10) and (11), let us introduce a Lyapunov-like function

$$W = \frac{1}{2} \sum_{i=1}^N \Bigg( \sum_{j \in \mathcal{N}_i^{\mathrm{sf}}(t)} \Upsilon_{ij}^{\mathrm{e}}(\|y_{ij}\|) + \sum_{j \in \mathcal{N}_i^{\mathrm{sz}}(t)} \Upsilon_{ij}^{\mathrm{c}}(\|y_{ij}\|) + y_i \sum_{j=1}^N G_{ij}(t) y_{ij} + \rho_i^T \rho_i \Bigg). \quad (13)$$

Consider the time interval $[t_0, t_1)$, one has $\Upsilon_{ij}^{\mathrm{e}} > 0$ from (10), $\Upsilon_{ij}^{\mathrm{c}} \geq 0$ from (11), and $\rho_i^T \rho_i \geq 0$. In addition, $\sum_{i=1}^N y_i \sum_{j=1}^N G_{ij}(t) y_{ij} = \sum_{i=1}^N y_i \sum_{j=1}^N L_{ij}(t) y_i = y^T (L(t) \otimes I_n) y \geq 0$ on account of the fact that $L(t) = L(t_0)$, $\mathcal{G}(t_0)$ is connected. Thus, one has that $W_0 = W(t_0) > 0$. Moreover, for $t \in [t_0, t_1)$, $G_{ij}(t)$ is fixed, one has

$$\dot{W} = \frac{1}{2} \sum_{i=1}^N \Bigg( \sum_{j \in \mathcal{N}_i^{\mathrm{sf}}(t)} \dot{\Upsilon}_{ij}^{\mathrm{e}}(\|y_{ij}\|) + \sum_{j \in \mathcal{N}_i^{\mathrm{sz}}(t)} \dot{\Upsilon}_{ij}^{\mathrm{c}}(\|y_{ij}\|) \Bigg)$$
$$+ \sum_{i=1}^N \dot{y}_i \sum_{j=1}^N L_{ij} y_j + \sum_{i=1}^N \rho_i^T \dot{\rho}_i$$
$$= \sum_{i=1}^N \sum_{j \in \mathcal{N}_i^{\mathrm{sf}}(t)} \dot{y}_i^T \nabla_{y_i} \Upsilon_{ij}^{\mathrm{e}}(\|y_{ij}\|) + \sum_{i=1}^N \dot{y}_i \sum_{j=1}^N L_{ij} y_j \quad (14)$$
$$+ \sum_{i=1}^N \sum_{j \in \mathcal{N}_i^{\mathrm{sz}}(t)} \dot{y}_i^T \nabla_{y_i} \Upsilon_{ij}^{\mathrm{c}}(\|y_{ij}\|) + \sum_{i=1}^N \rho_i^T \dot{\rho}_i$$
$$= -\rho^T (L(t_0) \otimes I_n) \rho. \qquad \square$$

Taking into account that $\mathcal{G}(t_0)$ is connected, one has $L(t_0) \geq 0$, which implies that $\dot{W} \leq 0$. Thus, $W(t) \leq W(t_0) \leq \mu_{\max}$, for $t \in [t_0, t_1)$. From (10), (11) and Remark 1, one has that $\Upsilon_{ij}^{\mathrm{e}}(\hat{r}_s) = \mu_1 > \mu_{\max}$, and $\Upsilon_{ij}^{\mathrm{c}}(\hat{d}_s) = \mu_2 > \mu_{\max}$, which yields that no collision appears during $[t_0, t_1)$, and no agent $j$ has left the set $\mathcal{N}_i^{\mathrm{sf}}$ for agent $i$. Hence, the network $\mathcal{G}(t)$ is still connected. Let us consider $t = t_1$, we assume that the number of new agents added in the set $\mathcal{N}_i^{\mathrm{sz}}$ is $k_i$ for agent $i$.

One has that $\sum_{i=1}^N k_i + \sum_{i=1}^N \mathrm{num}_i(\mathcal{N}_i^{\mathrm{sz}}) \leq N(N-1)$, and $\mathrm{num}_i(\mathcal{N}_i^{\mathrm{sz}})$ is the number of agents in $\mathcal{N}_i^{\mathrm{sz}}$. It yields that

$$W(t_1) \leq W(t_1^-) + \sum_{i=1}^N k_i \widetilde{\Upsilon} \leq W(t_0) + \sum_{i=1}^N k_i \widetilde{\Upsilon}$$
$$\leq \frac{1}{2} \sum_{i=1}^N \Bigg( \sum_{j \in \mathcal{N}_i^{\mathrm{f}}} \Upsilon_{ij}^{\mathrm{e}}(\|\hat{r}_s - \hat{\varepsilon}\|) +$$
$$+ y_i(t_0)^T \sum_{j=1}^N G_{ij}(t_0) y_{ij}(t_0) + \rho_i(t_0)^T \rho_i(t_0)$$
$$+ \sum_{j \in \mathcal{N}_i^{\mathrm{sz}}(t)} \Upsilon_{ij}^{\mathrm{c}}(\|y_{ij}\|) \Bigg) + \sum_{i=1}^N k_i \widetilde{\Upsilon}$$
$$< \mu_{\max}, \quad (15)$$

where $\widetilde{\Upsilon} = \frac{1}{2} \sum_{j \in \mathcal{N}_i^{\mathrm{sz}}} \Upsilon_{ij}^{\mathrm{c}}(\|\hat{d}_s - \hat{\varepsilon}\|)$. One can apply the above analysis for time intervals $[t_l, t_{l+1})$. The condition still holds that $\dot{W}(t) \leq 0$, and one has

$$W(t) \leq W(t_l) \leq \mu_{\max}, \quad (16)$$

which implies that there is no collision during $[t_l, t_{l+1})$, and no agent $j$ has left the set $\mathcal{N}_i^{\mathrm{sf}}$ for agent $i$. Hence, the graph $\mathcal{G}(t)$ is connected for $t \in [t_l, t_{l+1})$.

For the statement 3), let us assume that the edge set $\mathcal{E}(t)$ changes at $t_l$, $l = 0, 1, 2, \ldots$, and there is a time $\hat{t}_l$ such that the topology of $\mathcal{G}$ is fixed. For $t \in [\hat{t}_l, \infty)$, from the construction of $W$, one has that

$$\frac{1}{2} \sum_{i=1}^N y_i \sum_{j=1}^N G_{ij}(t) y_{ij} \leq \mu_{\max}, \quad \frac{1}{2} \sum_{i=1}^N \rho_i^T \rho_i \leq \mu_{\max}.$$

When the topology of $\mathcal{G}$ is fixed, one has that $G_{ij}$ is also fixed for $t \in [\hat{t}_l, \infty)$. On account of the symmetry of $G$, let $\lambda_{\max}$ be the largest eigenvalue of $G$, one has that

$$\frac{1}{2} y^T (L(\hat{t}_l) \otimes I_n) y \leq \frac{1}{2} \lambda_{\max} \|y\|^2 \leq \mu_{\max},$$

which yields that $\|y\| \leq \sqrt{\frac{2\mu_{\max}}{\lambda_{\max}}}$. Via similar arguments, one has that $\|\rho\| \leq \sqrt{2\mu_{\max}}$. Let us consider the set $\Xi = \{y \in \mathbb{R}^{Nn}, \rho \in \mathbb{R}^{Nn} | W(y, \rho) \leq \mu_{\max}, \|y\| \leq \sqrt{\frac{2\mu_{\max}}{\lambda_{\max}}}, \|\rho\| \leq \sqrt{2\mu_{\max}}\}$, which is a compact set. Now, let us study the largest invariant set in $\mathcal{I} = \{y \in \mathbb{R}^{Nn}, \rho \in \mathbb{R}^{Nn} | \dot{W} = 0\}$.

Based on (14), one has

$$\dot{W} = -\rho(L \otimes I_n)\rho = \frac{1}{2} \sum_{i \in \mathcal{N}, \, j \in \mathcal{N}_i^{\mathrm{s}}} G_{ij} \|\rho_i - \rho_j\|^2,$$

which implies that $\dot{W} = 0$ if and only if $\rho_1 = \cdots = \rho_N$. From LaSalle's invariance principle [17], it yields that all the trajectories started from $\Xi$ will eventually converge to $\mathcal{I}$, i.e., $\rho_1 = \cdots = \rho_N$.

For statement 4), consider the case of $t \geq \hat{t}_l$, one has

$\rho_i - \rho_j = 0$ for all $i, j \in \mathcal{N}$. Then, (12) can be rewritten as

$$u_i = -\sum_{j \in \mathcal{N}_i^{\text{sf}}(t)} \nabla_{y_i} \Upsilon_{ij}^{\text{e}}(\|y_{ij}\|) - \sum_{j \in \mathcal{N}_i^{\text{sz}}(t)} \nabla_{y_i} \Upsilon_{ij}^{\text{c}}(\|y_{ij}\|)$$
$$- \sum_{j \in \mathcal{N}_i^{\text{s}}(t)} G_{ij}(t) y_{ij},$$
$$= -\sum_{j \in \mathcal{N}_i^{\text{sf}}(t)} \frac{\partial \Upsilon_{ij}^{\text{e}}(\|y_{ij}\|)}{\partial \|y_{ij}\|} \cdot \frac{1}{\|y_{ij}\|} y_{ij} - \sum_{j \in \mathcal{N}_i^{\text{s}}(t)} G_{ij}(t) y_{ij}$$
$$- \sum_{j \in \mathcal{N}_i^{\text{sz}}(t)} \frac{\partial \Upsilon_{ij}^{\text{c}}(\|y_{ij}\|)}{\partial \|y_{ij}\|} \cdot \frac{1}{\|y_{ij}\|} y_{ij}.$$

From (10), one has that $\frac{\partial \Upsilon_{ij}^{\text{e}}(\|y_{ij}\|)}{\partial \|y_{ij}\|} \cdot \frac{1}{\|y_{ij}\|}$ is positive and bounded as $\|y_{ij}\| \to 0$, one has that $u_i = -(\tilde{L}(t) \otimes I_n + L(t) \otimes I_n)y$ with $\tilde{L}(t) \geq 0$ and $L(t) \geq 0$ as $t > \hat{t}_l$. From algebraic graph theory [1], it yields that $\lim_{t \to \infty} y = \text{span}(1_{Nn})$, i.e., $y_i - y_j = 0$, for all $i, j \in \mathcal{N}$. $\square$

*B. Computing SG-RMTC via Lyapunov-Like Barrier Functions*

In this subsection, a method based on SOS programming is proposed to enlarge the set $\mathcal{W}(c)$ by selecting fixed $\Upsilon_{ij}^{\text{e}}$ and fixed $\Upsilon_{ij}^{\text{c}}$, i.e., we aim at finding

$$\gamma = \sup \ c \qquad (17)$$

such that (10) and (11) hold. To increase the scalability of this method, we assume that $\Upsilon_{ij}^{\text{e}} = \Upsilon^{\text{e}}$ and $\Upsilon_{ij}^{\text{c}} = \Upsilon^{\text{c}}$.

To this end, we consider barrier functions in polynomial vector fields. It can be extended to non-polynomial or rational vector fields [5], which is outside the scope of this paper. First, let us introduce the Real Positivestellensatz, which provides a powerful tool to check the positivity of polynomials over semi-algebraic sets by exploiting the cone of SOS.

*Lemma 2 ([19]):* For polynomials $a_1, \ldots, a_m, b_1, \ldots, b_l$ and $p$, define a set

$$\mathcal{B} = \{x \in \mathbb{R}^n : a_i(x) = 0, \forall i = 1, \ldots, m, \\ b_i(x) \geq 0, \forall j = 1, \ldots, l\}. \qquad (18)$$

Let $\mathcal{B}$ be compact. Condition $\forall x \in \mathcal{B} : p(x) > 0$ can be established if

$$\begin{cases} \exists r_1, \ldots, r_m \in \mathcal{P}, \ s_1, \ldots, s_l \in \mathcal{P}^{\text{SOS}}, \\ p - \sum_{i=1}^m r_i a_i - \sum_{i=1}^l s_i b_i \in \mathcal{P}^{\text{SOS}}. \end{cases} \qquad (19)$$

*Remark 2:* Condition (19) turns to be a non-conservative condition if there is no degree bound for $s_i$, and if there is a polynomial $b$ in $\mathcal{B}$ such that $b^{-1}[0, \infty)$ is compact. $\square$

Based on the above result, a lower bound of $\gamma$ in (17) can be calculated by an SOS programming.

*Theorem 2:* Assume there exist functions $\Upsilon^{\text{e}}$ and $\Upsilon^{\text{c}}$ satisfying (10) and (11), respectively, and there exist polynomials $r_i(q) \in \mathcal{P}^{\text{SOS}}$, for all $i = 1, \ldots, h$, and a polynomial $s(q) \in \mathcal{P}^{\text{SOS}}$ such that $\bar{c}$ is the solution of the following optimization:

$$\bar{c} = \sup_{c, s} c$$
$$\text{s.t.} \begin{cases} -\psi(q, c, s(q), r_i(q)) \in \mathcal{P}^{\text{SOS}}, \\ \forall q \in \mathcal{W}(c) \setminus \{\bar{q}\}, \end{cases} \qquad (20)$$

where $\bar{q} = 0_{2N}$ is introduced in Section II, and

$$\psi(q, c, s(q), r_i(q)) = \dot{W}(q) + s(q)(c - W(q)) \\ + \sum_{i=1}^h r_i(q) w_i(q). \qquad (21)$$

Then, $\bar{c} \leq \gamma$.

*Proof:* Suppose (20) holds, one has that $-\psi(q, c, s(q), r_i(q))$ and $r_i(q)$ as well as $s(q)$ are SOS. From Lemma 2, it yields that

$$\dot{W}(q) < 0, \qquad (22)$$

for all $q$ in $\{x \in \mathbb{R}^{2N} : c - W(q) \geq 0\} \setminus \{\bar{q}\}$. Therefore, from (14) and the proof of Theorem 1, $\mathcal{W}(\bar{c})$ is an estimate of the SG-RMTC. Taking into account the definition of $\gamma$ in (17), it finally yields that $\bar{c}$ is a lower bound of $\gamma$, which completes this proof. $\square$

*Remark 3:* Theorem 2 transforms the condition of (14) to an SOS programming by using Lemma 2. It paves the way for generating more tractable methods by using LMIs. Along with Remark 2, the conservatism of above result relies on the degree of $s$ and $r_i$, and the relaxations of Lemma 2 [20]. $\square$

*C. Quasi-Convex Optimization via SMR*

The condition (20) of Theorem 2 is usually not easy to check since the product of $s(x)$ and $c$ makes it a bilinear inequality which is non-convex in nature. In this subsection, we will show how a generalized eigenvalue problem is obtained from the problem (20) by using the SMR technique. Specifically, for the class of polynomial $p_0(x) \in \mathcal{P}^{\text{SOS}}$, its SMR is as follows:

$$p_0(x) = (*)^T (\bar{P}_0 + L(\delta)) \phi(n, d_{p_0}), \qquad (23)$$

where $(*)^T AB$ is short for $B^T AB$ given in Section II, $\bar{P}_0$ denotes the SMR matrix of $p_0(x)$, $n$ is the number of variables, $d_{p_0}$ is the smallest integer not less than $\frac{\deg(p_0)}{2}$, i.e., $d_{p_0} = \lceil \frac{\deg(p_0)}{2} \rceil$, $\phi(n, d_{p_0}) \in \mathbb{R}^{l(n, d_{p_0})}$ is called the power vector including all monomials of degree less or equal to $d_{p_0}$, $L(\delta)$ is a parameterization of the space

$$\mathscr{L} = \{L(\delta) \in \mathbb{R}^{l(n, d_{p_0}) \times l(n, d_{p_0})} : L(\delta) = L^T(\delta), \\ (*)^T L(\delta) \phi(n, d_{p_0}) = 0\},$$

in which $\delta \in \mathbb{R}^{\vartheta(n, d_{p_0})}$ is a vector of free parameters. The functions $l(n, d_{p_0})$ and $\vartheta(n, d_{p_0})$ can be calculated as in [20]. For the purpose of clarity, an illustration is given:

*Example 1:* Given the polynomial $p_1(x) = 3x^4 + 4x^3 + 6x^2 + 7$, we have $d_{p_1} = 2$, $n = 1$ and $\phi(n, d_{p_1}) = (x^2, x^1, 1)^T$. Then, $p_1(x)$ can be written in (23) as:

$$\bar{P}_1 = \begin{pmatrix} 3 & 2 & 0 \\ 2 & 6 & 0 \\ 0 & 0 & 7 \end{pmatrix}, \ L(\delta) = \begin{pmatrix} 0 & 0 & -\delta \\ 0 & 2\delta & 0 \\ -\delta & 0 & 0 \end{pmatrix}.$$

Define $r(q) = (r_1(q), \ldots, r_h(q))^T$, $\xi(q) = \sum_{j=0}^{h} r_j(q)\omega_j(q)$, and let $\deg(\dot{W}) - \deg(W) \leq \deg(s)$, $\deg(\dot{W}) - \deg(\omega_j) \leq \deg(r_j)$, for all $j = 0, 1, \ldots, h$. From (23), we have the following expressions of SMR:

$$W(q) = (*)^T \overline{W} \phi(2N, d_w), \quad (24)$$
$$s(q) = (*)^T \overline{S} \phi(2N, d_s), \quad (25)$$
$$r_j(q) = (*)^T \overline{R}_j \phi(2N, d_{r_j}), \quad (26)$$
$$\psi(q) = (*)^T \overline{\Psi}(\delta, c, \overline{S})\phi(2N, d_\psi), \quad (27)$$

where $\delta \in \mathbb{R}^{\vartheta(2N, d_\psi)}$ is a vector of free parameters, $\overline{W} \in \mathbb{R}^{l(2N, d_w) \times l(2N, d_w)}$, $\overline{S} \in \mathbb{R}^{l(2N, d_s) \times l(2N, d_s)}$ and $\overline{\Psi}(\delta, c, \overline{S}, \Xi) \in \mathbb{R}^{l(2N, d_\psi) \times l(2N, d_\psi)}$ are symmetric matrices. Let $\overline{D}(\delta)$, $\Xi$, $\Lambda_1(\overline{S})$ and $\Lambda_2(\overline{S})$ be SMR matrices of $\dot{W}(q)$, $\xi(q)$, $s(q)$ and $W(q)s(q)$, respectively, with respect to the power vector $\phi(2N, d_\psi)$. From (21), it yields

$$\Psi(\delta, c, \overline{S}, \Xi) = \overline{D}(\delta) + \Xi(\overline{R}_j) + c\Lambda_1(\overline{S}) - \Lambda_2(\overline{S}),$$

where $\delta \in \mathbb{R}^{\vartheta(2N, d_\psi)}$ is a vector of free parameters. The following result transforms the condition (20) into a generalized eigenvalue problem (GEVP).

*Theorem 3:* For given positive scalars $\sigma_1$, $\sigma_2$, and a selected polynomial $W(q, \Upsilon^e, \Upsilon^c) = (*)^T \overline{W} \phi(2N, d_w)$ with chosen $\Upsilon^e, \Upsilon^c$ fulfilling (10) and (11), respectively, the polynomial $\varsigma(q) = \sigma_1 s(q) + \sigma_2 W(q)s(q) = (*)^T \Lambda(\overline{S})\phi(2N, d_\psi)$, the lower bound of $\gamma$ can be obtained by

$$\tilde{\gamma} = -\frac{\tilde{e}}{\sigma_1 + \sigma_2 \tilde{e}}, \quad (28)$$

where $\tilde{e}$ is the solution of the GEVP

$$\tilde{e} = \inf_{\delta, e, \overline{S}} e$$
$$\text{s.t.} \begin{cases} \sigma_1 + \sigma_2 e > 0, \\ \overline{S} > 0, \\ e\Lambda(\overline{S}) > \overline{D}(\delta) - \Xi(\overline{R}_j) - \Lambda_2(\overline{S}). \end{cases} \quad (29)$$

*Proof:* In this proof, we first show that 1) (29) is a GEVP. Then, we demonstrate 2) (28) is the lower bound of $\tilde{\gamma}$.

First, we aim to prove the optimization (29) is a GEVP: From [21], we have $\Lambda > 0$ on the condition that $\overline{W} > 0$ and $\overline{S} > 0$, which makes (29) a GEVP.

Second, we are trying to show that $\tilde{\gamma}$ in (28) is the lower bound of $\tilde{\gamma}$: Based on the last inequality of (29), we have

$$\tilde{\Phi}(\delta, c, \overline{S}) = \overline{D}(\delta) - \Xi(\overline{R}_j) - e\Lambda(\overline{S}) - \Lambda_2(\overline{S}) < 0.$$

Considering (27) and

$$\tilde{\psi}(q, c, s(q), r(q)) = \dot{W}(q) - \xi(r(q), q) - W(q)s(q) - e(\sigma_1 + \sigma_2 W(q))s(q),$$

one can rewrite $\tilde{\psi}(q, c, s(q), r(q))$ into:

$$\tilde{\psi}(q, c, s(q), r(q)) = \tilde{\psi}(q, \frac{-e}{\sigma_1 + \sigma_2 e}, (\sigma_1 + \sigma_2 e)s(q), r(q)).$$

Notice that $-e/(\sigma_1 + \sigma_2 e)$ is a monotonically decreasing function which maps from the range $(-(\sigma_1/\sigma_2), 0]$ into the range $[0, +\infty)$. Thus, (28) gives the lower bound of $\tilde{\gamma}$. □

For more details of the GEVP, please see the book [21].

### D. The Optimal Lyapunov-Like Barrier Functions

In this subsection, strategies for finding the optimal $\Upsilon^e(q)$ and $\Upsilon^c(q)$ are proposed. First, let us recall that $\rho$ in Problem 1 is a user-selected measure which is often chosen as

$$\rho(\mathcal{W}(\gamma)) = \text{vol}(\mathcal{W}(\gamma)),$$

where $\text{vol}(\mathcal{W}(\gamma))$ denotes the volume of $\mathcal{W}(\gamma)$, and $\gamma$ is introduced in (17). This paves a way to pursue the optimal $W(q, \Upsilon^e, \Upsilon^c)$ via maximizing the volume of $\mathcal{W}(\gamma)$. However, $\text{vol}(\mathcal{W}(\gamma))$ is highly non-convex, which makes (17) a non-convex optimization. To solve this problem, a typical method is to approximate $\text{vol}(\mathcal{W}(\gamma))$ by introducing

$$\eta = \max \frac{\gamma^n}{\det(\overline{W}(\overline{\Upsilon}^e, \overline{\Upsilon}^c))}, \quad \text{vol}(\mathcal{W}(\gamma)) \propto \eta, \quad (30)$$

where $\overline{W}$ is the SMR matrix of $W(x)$ in (24), $\overline{\Upsilon}^e$ and $\overline{\Upsilon}^c$ are SMR matrices of $\Upsilon^e$ and $\Upsilon^c$ with

$$\begin{aligned} \Upsilon^e(q) &= (*)^T \overline{\Upsilon}^e \phi(2N, d_w), \\ \Upsilon^c(q) &= (*)^T \overline{\Upsilon}^c \phi(2N, d_w), \end{aligned} \quad (31)$$

and $\text{vol}(\mathcal{W}(\gamma))$ is proportional to $\omega$. Then, a linear approximation of $\text{vol}(\mathcal{W}(\gamma))$ can be provided as

$$\text{vol}(\mathcal{W}(\gamma)) \approx \frac{\gamma}{\text{trace}(\overline{W})}. \quad (32)$$

The underlying idea is to minimize $\text{trace}(\overline{W})$ instead of the non-convex objective with $\det(\overline{W})$. Thus, a strategy is given for searching the optimal $\Upsilon^e$ and $\Upsilon^c$:

Assume that there exist $s \in \mathcal{P}^{\text{SOS}}$ and $r_j \in \mathcal{P}^{\text{SOS}}$, for all $j = 1, \ldots, h$, such that

$$\zeta = \inf_{\overline{\Upsilon}^e, \overline{\Upsilon}^c} \text{trace}(\overline{W}(\overline{\Upsilon}^e, \overline{\Upsilon}^c))$$
$$\text{s.t.} \begin{cases} W(\overline{\Upsilon}^e, \overline{\Upsilon}^c, q) \in \mathcal{P}^{\text{SOS}}, \\ (10) - (11) \text{ hold}, \\ -\psi(q, \overline{\Upsilon}^e, \overline{\Upsilon}^c, s, r) \in \mathcal{P}^{\text{SOS}}. \end{cases} \quad (33)$$

Then, $\kappa_1 = \frac{\gamma}{\zeta}$ is an under-estimate of $\rho$.

The condition of (33) could be transformed to SOS programmings. Specifically, from Lemma 2, it is not difficult to obtain that (10) holds if there exist $z \in \mathbb{R}$, $\Upsilon^e(\overline{\Upsilon}^e, z) \in \mathcal{P}^{\text{SOS}}$, $\tilde{s}_1(z) \in \mathcal{P}^{\text{SOS}}$, and $\tilde{s}_2(z) \in \mathcal{P}^{\text{SOS}}$, such that

$$\begin{cases} \Upsilon^e(\overline{\Upsilon}^e, \hat{r}_s) = \mu_1, r^e(\overline{\Upsilon}^e, z) \in \mathcal{P}^{\text{SOS}}, \\ -d^e(\overline{\Upsilon}^e, z) - \tilde{s}_1 z - \tilde{s}_2(\hat{r}_s - z) \in \mathcal{P}^{\text{SOS}}. \end{cases} \quad (34)$$

where $d^e(\overline{\Upsilon}^e, z) = \frac{\partial \Upsilon^e}{\partial z}$ and $r^e(\overline{\Upsilon}^e, z) = \frac{\partial \Upsilon^e}{\partial z} \cdot \frac{1}{z}$. Moreover, (11) holds if there exist $z \in \mathbb{R}$, $\Upsilon^c(\overline{\Upsilon}^c, z) \in \mathcal{P}^{\text{SOS}}$, and $\tilde{s}_3(z) \in \mathcal{P}^{\text{SOS}}$ such that

$$\begin{cases} \Upsilon^c(\overline{\Upsilon}^c, \hat{d}_s) = \mu_2, \\ -d^c(\overline{\Upsilon}^c, z) - \tilde{s}_1 z - \tilde{s}_2(\hat{r}_s - z) \in \mathcal{P}^{\text{SOS}}. \end{cases} \quad (35)$$

where $d^c(\overline{\Upsilon}^c, z) = \frac{\partial \Upsilon^c}{\partial z}$. Then, (33) can be transformed to tractable conditions as follows:

*Proposition 1:* Assume that there exist $s \in \mathcal{P}^{\text{SOS}}$ and local SOS polynomials $\Upsilon^e$, $\Upsilon^c$, $\tilde{s}_1(z)$, $\tilde{s}_2(z)$, $\tilde{s}_3(z)$, $r_j$, $\forall j =$

$0, 1, \ldots, h$, such that

$$\zeta = \inf_{\bar{\Upsilon}^e, \bar{\Upsilon}^c, c} \text{trace}(\overline{W}(\bar{\Upsilon}^e, \bar{\Upsilon}^c))$$
$$\text{s.t.} \begin{cases} W(\bar{\Upsilon}^e, \bar{\Upsilon}^c, q) \in \mathcal{P}^{\text{SOS}}, \\ (34) - (35) \text{ hold}, \\ -\psi(q, \bar{\Upsilon}^e, \bar{\Upsilon}^c, s, r, c) \in \mathcal{P}^{\text{SOS}}. \end{cases} \quad (36)$$

Then, $\kappa_2 = \frac{\gamma}{\zeta}$ is an under-estimate of $\rho$.

Observe that the last constraint of (36) can be rewritten as $-\tilde{w}(q) - s(q)(c - W(\bar{\Upsilon}^e, \bar{\Upsilon}^c)) + \sum_{i=1}^h r_i(q) w_i(q)$ where $\tilde{w}(q) = -\rho^T (L(t) \otimes I_n) \rho$ from (14). In order to cope with this, one useful way is by iterating among $s(q)$ and $c$ (using the technique for the fixed Lypuanov-like barrier functions shown in Section III.B-C) and $\bar{\Upsilon}^e, \bar{\Upsilon}^c$, which returns an iterative LMIs problem and it can be solved by existing delicate softwares, as illustrated in the following section.

## IV. SIMULATIONS

To illustrate the proposed approach, a numerical example of smart cars platooning is provided. We execute the computation using MATLAB R2017a on a desktop with a 16GB DDR3 RAM and an Intel Xeon E3-1245 processor (3.4 GHz). The MATLAB toolbox SeDuMi is used for solving semi-definite problems.

In this example, an implementation with autonomous driving is considered. The safe platooning of cars can be achieved if the proposed method ensures the multi-objective coordination of smart cars without entering the unsafe areas, which are represented as construction areas and a broken yellow car as shown in Fig. 2. Each smart car (red) is assumed to be an agent, whose model is set up with the following parameters: $r_a = 0.75$, $r_s = 11$, $r_z = 3.5$, $r_c = 1.25 r_a$, $d_s = 2 r_c$, and $\epsilon = 0.1$.

The unsafe area $\Omega = \Omega_1 \cup \Omega_2 \cup \Omega_3 \cup \Omega_4 \cup \Omega_5$ given by (4) is expressed by following polynomial inequalities,

$$\begin{aligned} \Omega_1 &= \{x \in \mathbb{R}^2 | (x_i(1) - 8)^2 + (x_i(2) - 4)^2 - 4 < 0\}, \\ \Omega_2 &= \{x \in \mathbb{R}^2 | x_i(1) > 7, \ x_i(2) < -2\}, \\ \Omega_3 &= \{x \in \mathbb{R}^2 | x_i(1) < 0, \ x_i(2) > 2\}, \\ \Omega_4 &= \{x \in \mathbb{R}^2 | x_i(2) < -6\}, \\ \Omega_5 &= \{x \in \mathbb{R}^2 | x_i(2) > 6\}, \end{aligned}$$

where $\Omega_1$ encodes the area of the broken car, $\Omega_2$ and $\Omega_3$ describe the areas under construction, $\Omega_4$ and $\Omega_5$ describe the boundaries of road.

First, let us check whether the multi-objective coordination is achieved by the proposed controller (12). From Fig. 3, we could see that the platooning of smart cars is obtained and the differences of velocities converge to 0, and these smart cars are kept away from the unsafe areas. In addition, for the connectivity maintenance, distributed controllers preserve the edges $(A_1, A_2)$ and $(A_2, A_3)$, and allow break of the edge $(A_1, A_3)$ as system evolves, which ensures the connectivity of the whole network. Demonstrated by Fig. 4, the collision avoidance amongst smart cars is also guaranteed. As we could see from Fig. 3, the car 3 moves backward first to avoid collision with car 2 when it is merging in the middle lane.

Then, let us consider fixed Lyapunov-like barrier functions with $\Upsilon^e = c_1(\|y_{ij}\|)^4$ and $\Upsilon^c = c_2(\|y_{ij}\|^2 - \tilde{r}_z^2)^2$, where $c_1 = \frac{\mu_1}{\tilde{r}_s^4}$ and $c_2 = \frac{\mu_2}{\tilde{d}_s - \tilde{r}_z^2}$. Then, we compute the optimal Lyapunov-like barrier function by using Theorem 3 and Proposition 1, and one has $\zeta = 16.3245$. The computational results are shown in Fig. 5, from which the estimate of SG-RMTC is significantly enlarged by using the optimal Lyapunov-like barrier function compared to the method of fixed Lyapunov-like barrier functions.

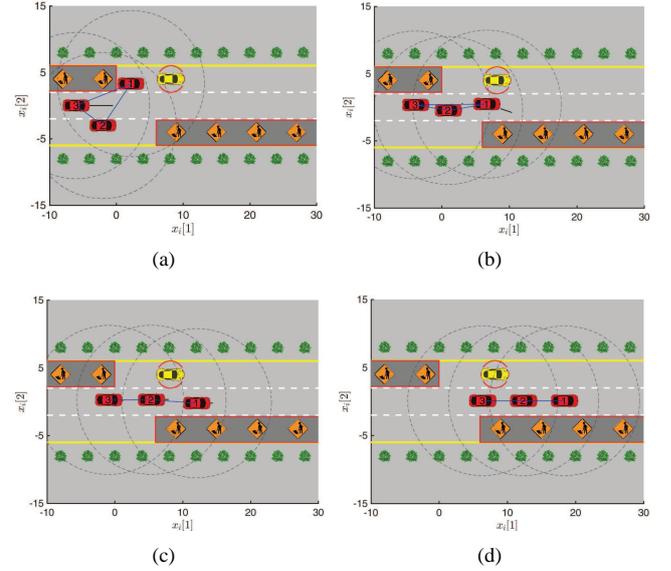

Fig. 2. The motion of cars and the set of edges for t = 0, 1, 3, 9, respectively.

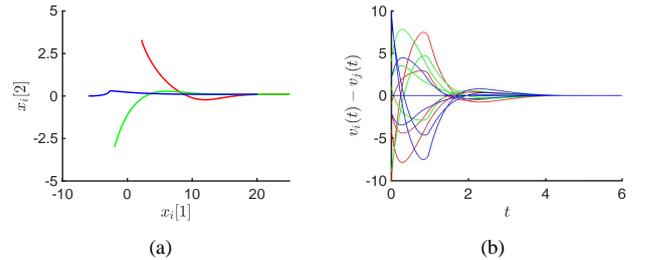

Fig. 3. The trajectories of agents and the differences of velocities.

TABLE I

THE COMPUTATIONAL TIME $t_c$ [sec] FOR DIFFERENT NUMBERS OF ITERATIONS $n_t$, AND DEGREES OF BARRIER FUNCTIONS $d_b$.

|  | $d_b = 2$ | | | $d_b = 4$ | | |
| --- | --- | --- | --- | --- | --- | --- |
|  | $n_t=5$ | $n_t=10$ | $n_t=20$ | $n_t=5$ | $n_t=10$ | $n_t=20$ |
| $t_c$ | 17.52 | 29.63 | 68.51 | 112.3 | 214.5 | 407.2 |

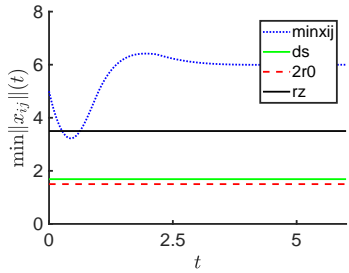

Fig. 4. The minimal distance between smart cars.

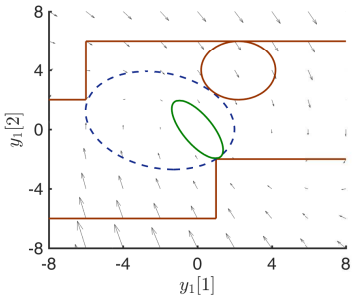

Fig. 5. Computational results of the estimates of SG-RMTC for car 1. The solid red lines depict the boundaries of unsafe sets; the solid green line represents the estimate via a fixed Lypaunov-like barrier function with degree 4; the dashed blue line represents the estimate via the optimal Lyapunov-like barrier function with degree 2.

Note that static unsafe sets are considered in this case, this method is flexible to extend to the situation with moving unsafe sets by considering additional barrier terms [22].

## V. Conclusion and Discussion

Multi-task coordination of multi-agent systems is considered, with objectives including convergence, collision avoidance, connectivity maintenance, and safety assurance. The problem of estimating the safety guaranteed region of multi-task coordination (SG-RMTC) is formulated. To cope with this problem, the sublevel set of Lyapunov-like barrier function is used, and a systematic way of constructing such kind of functions is proposed via Sum-of-Squares (SOS) programming and Square Matrix Representation (SMR). By searching the optimal Lyapunov-like barrier function, the best estimate of SG-RMTC can be obtained.

Future efforts will be devoted to designing a less-conservative convex approach for approximating the SG-RMTC, e.g., using the *moment theory* [23], enlarging the lower bound of $\mu$ via *rational ploynomial Lypunov-like barrier functions*, and combining *multiple sublevel sets* of Lypunov-like barrier functions. In addition, we are interested to compare this approach with other stability verification methods, like the contraction theory [24].